\documentclass[a4paper,11pt]{article}
\usepackage{pifont}
\date{}

\usepackage{makeidx}
\usepackage[centertags]{amsmath}
\usepackage{amsfonts}
\usepackage{amssymb}
\usepackage{amsthm}
\usepackage{epsfig}
\usepackage{color}
\makeatletter \@addtoreset{equation}{section} \makeatother

\begin{document}
\title{\bf Nonplanar Periodic Solutions for Spatial Restricted N+1-Body Problems  \footnote{Supported
by National Natural Science Foundation of China.}}

\author{{ Fengying Li\ and \ Shiqing Zhang\ and \ Xiaoxiao Zhao}\\
{\small Yangtze Center of Mathematics and College of Mathematics, Sichuan University,}\\
{\small Chengdu 610064, People's Republic of China}} \maketitle
\begin{quote}

{\bf Abstract:} We use variational minimizing methods to study
spatial restricted N+1-body problems with a zero mass moving on
the vertical axis of the moving plane for N equal masses. We prove
that the minimizer of the Lagrangian action on the anti-T/2 or odd
symmetric loop space must be a non-planar periodic solution for
any $N\geq2$.


2000 AMS Subject Classification 70F07, 34C25, 58E30
\end{quote}

\section{Introduction and Main Result}
\ \ \ \ \
    Spatial restricted 3-body model was studied by Sitnikov
\cite{r5}. Mathlouthis \cite{r3} etc. studied the periodic
solutions for the spatial circular restricted 3-body problems by
minimax variational methods.

    In this paper, we study spatial circular restricted N+1-body problems
with a zero mass moving on the vertical axis of the moving plane for
N equal masses. Suppose point masses $m_{1}=\cdots=m_{N}=1$ move
centered at the center of masses on a circular orbit. The motion for
the zero mass is governed by the gravitational forces of $m_{1},
\cdots, m_{N}$. Let $\rho_{j}=e^{\sqrt{-1}\frac{2\pi j}{N}}$  and
\begin{equation}\label{e1}
q_{1}(t)=re^{\sqrt{-1}2\pi t}\rho_{1}, \cdots, \
q_{j}(t)=\rho_{j}q_{1}(t), \cdots, \ q_{N}(t)=re^{\sqrt{-1}2\pi t }
\end{equation}
satisfy the Newtonian equations:
\begin{equation}\label{e2}
m_{i}\ddot{q_{i}}=\frac{\partial U}{\partial q_{i}},\ \ \ \
i=1,\cdots,N,
\end{equation}
where
\begin{equation}\label{e3}
U=\sum\limits_{1\leq i< j\leq N}\frac{m_{i}m_{j} }{| q_{i}-q_{j}|}.
\end{equation}
The orbit $q(t)=(0,0,z(t))\in R^{3}$ for zero mass satisfies the
following equation
\begin{equation}\label{e8}
\ddot{q}=\sum\limits_{i=1}^{N}\frac{m_{i}(q_{i}-q) }{|
q_{i}-q|^{3}}.
\end{equation}
Define

\begin{equation}
f(q)=\int_{0}^{1}\big[\frac{1}{2}|\dot{q}|^{2}+\sum\limits_{i=1}^{N}\frac{1
}{| q-q_{i}|}\big]dt,\ \ \ \ q\in \Lambda_{i},
\end{equation}
then
\begin{equation}\label{e4}
f(q)=\int_{0}^{1}\big[\frac{1}{2}|z'|^{2}+\frac{N}{\sqrt{r^{2}+z^{2}}}\big]dt\triangleq
f(z),\ \ \ \ q\in \Lambda_{i},
\end{equation}
where
\begin{equation*}
\Lambda_{1}=\left\{
\begin{array}{c}
q(t)=(0,0,z(t))|z(t)\in W^{1,2}(R/Z,R) \ \ \ \ \ \ \ \ \ \ \ \ \ \ \ \ \ \ \ \ \\
z(t+\frac{1}{2})=-z(t),\ q(t)\neq q_{i}(t),\ \forall t \in R,
i=1,2,\cdots,N
\end{array}
\right\},
\end{equation*}
\begin{equation*}
\Lambda_{2}=\left\{
\begin{array}{c}
q(t)=(0,0,z(t))|z(t)\in W^{1,2}(R/Z,R)  \\
q(-t)=-q(t)\ \ \ \ \ \ \ \ \ \ \ \ \ \ \ \ \ \ \ \ \ \ \ \ \ \ \ \ \
\ \
\end{array}
\right\},\ \ \ \ \ \ \ \ \ \ \ \ \ \ \ \ \ \ \ \ \
\end{equation*}

\begin{equation*}
W^{1,2}(R/Z,R)=\left\{x(t)\bigg|
\begin{array}{c}
x(t),\dot{x}(t)\in L^{2}(R,R) \\
x(t+1)=x(t)\ \ \ \ \ \ \ \
\end{array}
\right\}.
\end{equation*}
Notice that the symmetry in $\Lambda_1$ is related with Italian
symmetry [1].

In this paper,our main result is the following:

\vspace{0.4cm}\textbf{Theorem 1.1} \ The minimizer of $f(q)$ on
the closure $\overline{\Lambda}_{i}$ of $\Lambda_{i}$(i=1,2) is a
nonplanar and noncollision periodic solution.

\section{Proof of Theorem 1.1}
We define the inner product and equivalent norm of $W^{1,2}(R/Z,R)$:
\begin{equation}
<u,v>=\int_{0}^{1}(uv+u'\cdot v')dt,\ \ \ \ \ \ \ \ \ \ \ \ \ \
\end{equation}
\begin{equation}
\begin{aligned}
\|u\|&=\Big[\int_{0}^{1}|u|^{2}dt\Big]^{\frac{1}{2}}+\Big[\int_{0}^{1}
|u'|^{2}dt\Big]^{\frac{1}{2}}\\
&\cong\Big[\int_{0}^{1}|u'|^{2}dt\Big]^{\frac{1}{2}}+|u(0)|.
\end{aligned}
\end{equation}

\vspace{0.4cm}{\textbf{Lemma 2.1}}(Palais's Symmetry
Principle([4]))\ Let $\sigma$ be an orthogonal representation of a
finite or compact group $G$ in the real Hilbert space $H$ such that
for $\forall\sigma\in G, f(\sigma\cdot x)=f(x)$, where
$f:H\rightarrow R$.

Let $S=\{x\in H|\sigma\cdot x=x,\ \forall \sigma\in G\}$. Then the
critical point of $f$ in $S$ is also a critical point of $f$ in $H$.

By Palais's Symmetry Principle, we know that the critical point of
$f(q)$ in $\overline{\Lambda}_{i}$ is a noncollission periodic
solution of Newtonian equation (\ref{e8}).

In order to prove Theorem 1.1, we need

\vspace{0.4cm}{\textbf{Lemma 2.2}}([6]) Let $X$ be a reflexive
Banach space, $S$ be a weakly closed subset of $X$, $f:S\rightarrow
R\cup +\infty,\ \ f\not\equiv +\infty$ is weakly lower
semi-continuous and coercive($f(x)\rightarrow +\infty$ as
$\|x\|\rightarrow +\infty$), then $f$ attains its infimum on $S$.

\vspace{0.4cm}\textbf{Lemma 2.3}(Poincare-Wirtinger Inequality)\ \
Let $q\in W^{1,2}(R/Z,R^{N})$ and $\int_{0}^{T}q(t)dt=0$, then
\begin{equation}
\int_{0}^{T}|\dot{q}(t)|^{2}dt\geq\Big(\frac{2\pi}{T}\Big)^{2}\int_{0}^{T}|q(t)|^{2}dt.
\end{equation}

\vspace{0.4cm}{\textbf{Lemma 2.4}}\ \ $f(q)$ in (\ref{e4}) attains
its infimum on $\bar{\Lambda}_{1}=\Lambda_{1}$ or
$\bar{\Lambda}_{2}=\Lambda_{2}$.

\vspace{0.4cm}{\textbf{Proof.}}\ \ By Lemma 2.2 and Lemma 2.3, it is
easy to prove Lemma 2.4.

\vspace{0.4cm}{\textbf{Lemma 2.5}}(Jacobi's Necessary Condition[2])\
\ If the critical point $u=\tilde{u}(t)$ corresponds to a minimum of
the functional $\int_{a}^{b}F(t,u(t),u'(t))dt$ and if $F_{u'u'}>0$
along this critical point, then the open interval $(a,b)$ contains
no points conjugate to $a$, that is, for $\forall c\in(a,b)$, the
following boundary value problem:
\begin{equation}
\label{} \left\{\begin{array}{ll}
-\frac{d}{dt}(Ph')+Qh=0,& \\
h(a)=0,\ \ h(c)=0,
\end{array}\right.
\end{equation}
has only the trivial solution $h(t)\equiv0,\ \forall t\in(a,c)$,
where
\begin{equation}
P=\frac{1}{2}F_{u'u'}|_{u=\tilde{u}},\ \ \ \ \ \ \ \ \ \ \ \ \
\end{equation}
\begin{equation}
Q=\frac{1}{2}(F_{uu}-\frac{d}{dt}F_{uu'})|_{u=\tilde{u}}.
\end{equation}

\vspace{0.4cm}{\textbf{Lemma 2.6}}\ \ The radius $r$ for the moving
orbit of N equal masses is
\begin{equation*}
r=\Big(\frac{1}{4\pi}\Big)^{\frac{2}{3}}\Big[\sum\limits_{1\leq
j\leq N-1}csc(\frac{\pi}{N}j)\Big]^{\frac{1}{3}}.
\end{equation*}

\vspace{0.4cm}{\textbf{Proof.}}\ \ By (\ref{e1})-(\ref{e3}), we have
\begin{equation}\label{e5}
\ddot{q}_{N}=\sum\limits_{j\neq
N}\frac{q_{j}-q_{N}}{|q_{j}-q_{N}|^{3}},
\end{equation}
Substituting (\ref{e1}) into (\ref{e5}), we have
\begin{equation}
-4\pi^{2}=\sum\limits_{j\neq
N}\frac{\rho_{j}-\rho_{N}}{r^{3}|\rho_{j}-\rho_{N}|^{3}}
\end{equation}
\begin{equation}
\begin{aligned}
4\pi^{2}r^{3}&=\sum\limits_{j\neq N}\frac{1-\rho_{j}}{|1-\rho_{j}|^{3}} \\
&=\frac{1}{4}\sum\limits_{1\leq j\leq N-1}csc(\frac{\pi}{N}j)
\end{aligned}
\end{equation}
Then
\begin{equation}
r^{3}=\frac{1}{16\pi^{2}}\sum\limits_{1\leq j\leq
N-1}csc(\frac{\pi}{N}j).
\end{equation}
Therefore
\begin{equation}
r=\Big(\frac{1}{4\pi}\Big)^{\frac{2}{3}}\Big[\sum\limits_{1\leq
j\leq N-1}csc(\frac{\pi}{N}j)\Big]^{\frac{1}{3}}.
\end{equation}

\vspace{0.4cm}{\textbf{Lemma 2.7}}([8])\ \ $\sum\limits_{j=1}^{
N-1}csc(\frac{\pi}{N}j)=\frac{4}{N}$.

For the functional (\ref{e4}), let
\begin{equation*}
F(z,z')=\frac{1}{2}|z'|^{2}+\frac{N}{\sqrt{r^{2}+z^{2}}}.
\end{equation*}
Then the second variation of (\ref{e4}) in the neighborhood of $z=0$
is given by
\begin{equation}\label{e6}
\int_{0}^{1}(Ph'^{2}+Qh^{2})dt,\ \ \ \ \ \ \ \ \ \ \ \ \ \ \ \ \ \
\end{equation}
where
\begin{equation}
P=\frac{1}{2}F_{z'z'}|_{z=0}=\frac{1}{2},\ \ \ \ \ \ \ \ \ \ \ \ \ \
\ \ \ \ \
\end{equation}
\begin{equation}
Q=\frac{1}{2}(F_{zz}-\frac{d}{dt}F_{zz'})|_{z=0}=-\frac{N}{2r^{3}}.
\end{equation}
The Euler equation of (\ref{e6}) is called the Jacobi equation of
the original functional (\ref{e4}), which is
\begin{equation}
-\frac{d}{dt}(Ph'^{2})+Qh=0,\ \ \ \ \ \ \ \ \ \ \ \ \
\end{equation}
That is,
\begin{equation}\label{e7}
h''+\frac{N}{r^{3}}h=0.\ \ \ \ \ \ \ \ \ \ \ \ \
\end{equation}
Next, we study the solution of (\ref{e7}) with initial values
$h(0)=0,\ h'(0)=1$. It is easy to get
\begin{equation}
h(t)=\sqrt{\frac{r^{3}}{N}}\cdot sin\sqrt{\frac{N}{r^{3}}}t,
\end{equation}
which is not identically zero on $[0,\frac{1}{2}]$, but we will
prove $h(\frac{1}{2})=0$, and $h(c)=0$ for some
$c\in(0,\frac{1}{2})$. Notice that
\begin{equation}
\sqrt{\frac{N}{r^{3}}}=\sqrt{N}4\pi\Big(\sum\limits_{j\neq
N}csc\frac{\pi}{N}j\Big)^{-\frac{1}{2}}
\end{equation}
Hence
\begin{equation}
\begin{aligned}
\frac{1}{2}\sqrt{\frac{N}{r^{3}}}&=\sqrt{N}\Big(\sum\limits_{j\neq
N}csc\frac{\pi}{N}j\Big)^{-\frac{1}{2}}\cdot 2\pi \\
&=\sqrt{N}\Big(\frac{4}{N}\Big)^{-\frac{1}{2}}\cdot 2\pi \\
&=N\pi.
\end{aligned}
\end{equation}
So
\begin{equation}
h(\frac{1}{2})=0. \ \ \ \ \ \ \ \ \ \ \ \ \ \ \ \ \ \ \ \ \ \ \ \ \
\ \ \ \ \ \ \ \ \ \ \ \ \ \ \ \ \ \ \ \ \ \ \ \ \ \ \ \ \ \ \ \ \ \
\ \ \ \ \ \ \ \ \ \ \ \ \ \ \ \ \ \ \
\end{equation}
Given $N\geq2$, choose $0<c=\frac{1}{2N}<\frac{1}{2}$ such that
$2Nc=1$, then
\begin{equation}
\sqrt{\frac{N}{r^{3}}}c=2N\pi c=\pi
\end{equation}
Therefore
\begin{equation}
sin\sqrt{\frac{N}{r^{3}}}c=sin\pi=0.
\end{equation}
Hence $q(t)=(0,0,0)$ is not a local minimum for $f(q)$ on
$\bar{\Lambda}_{i}=\Lambda_{i}(i=1,2)$. So the minimizers of $f(q)$
on $\Lambda_{i}$ are not always at the center of masses, they must
oscillate periodically on the vertical axis, that is, the minimizers
are not always co-planar, hence we get the non-planar periodic
solutions.

\textbf{Acknowledgements} The authors sincerely thank the referee
for his/her valuable comments and  helpful suggestions.

\end{document}